\documentclass[twocolumn,prb,preprintnumbers,prb,amsmath,superscriptaddress,citeautoscript]{revtex4}
\usepackage{graphicx}
\usepackage{dcolumn}
\usepackage{bm}% bold math
\usepackage{natbib}
\begin{document}

%\preprint{APS/123-QED}

\title{Fabrication of micro-mirrors with pyramidal shape using anisotropic\\ etching of silicon}% Force line breaks with \\

\author{Z. Moktadir$^*$}
 \affiliation{School of Electronics and Computer Science,
Southampton University, Southampton, SO17 1BJ, United Kingdom}%Lines break automatically or can be forced with \\
 \email{zm@ecs.soton.ac.uk}
 
\author{C. Gollasch}
\affiliation{School of Electronics and Computer Science,
Southampton University, Southampton, SO17 1BJ, United Kingdom}

\author{E. Koukharenko}
\affiliation{School of Electronics and Computer Science,
Southampton University, Southampton, SO17 1BJ, United Kingdom}

\author{M. Kraft}
\affiliation{School of Electronics and Computer Science,
Southampton University, Southampton, SO17 1BJ, United Kingdom}

\author{G. Vijaya Prakash}
\affiliation{School of Physics and Astronomy, University of
Southampton, Southampton, SO17 1BJ, United Kingdom}

\author{J.J.
Baumberg} \affiliation{School of Physics and Astronomy, University
of Southampton, Southampton, SO17 1BJ, United Kingdom}

\author{M. Trupke}
\affiliation{Blackett Laboratory, Imperial College, Prince Consort
Road, London, United Kingdom}
\author{S. Eriksson}
\affiliation{Blackett Laboratory, Imperial College, Prince Consort
Road, London, United Kingdom}
\author{E.A. Hinds} \affiliation{Blackett Laboratory, Imperial College,
Prince Consort Road, London, United Kingdom}


\begin{abstract}
Gold micro-mirrors have been formed in silicon in an inverted
pyramidal shape. The pyramidal structures are created in the
$(100)$ surface of a silicon wafer by anisotropic etching in
potassium hydroxide. High quality micro-mirrors are then formed by
sputtering gold onto the smooth silicon (111) faces of the
pyramids. These mirrors show great promise as high quality optical
devices suitable for integration into MOEMS systems.
\end{abstract}

%\pacs{}% PACS, the Physics and Astronomy
\maketitle

The miniaturization of optical components leads to
 higher packaging density and increased speed of devices that manipulate
light.  This is part of the vast field of Microsystems technology,
designated by Micro-Opto-Electro-Mechanical Systems (MOEMS), in
which electronic, mechanical, and optical devices are integrated
on the micron scale. Mirrors are fundamental components of most
optical systems. A repeatable technique for the fabrication of
high-quality micro-mirrrors is therefore an essential step in the
advancement of this field. For example, if the surface quality of
the mirrors were high enough, one could contemplate using them to
make miniature optical resonators, which would have significant
applications ranging from narrow-band filters in optical
communications \cite{Djordjev} to the enhancement or suppression
of spontaneous emission in cavity QED devices \cite{Kimble}. At
present there is no simple MOEMS technique to make small mirrors.
Small optical cavities have been made using silica and quartz
microspheres (see for example ref. [3]), and Coyle {\it et
al.}\cite{Coyle} have reported spherical metallic mirrors
fabricated using templated self-assembly, but neither method is
ideally suited for integration with MOEMS technology.

\begin{figure}[!htb]
\begin{center}
\begin{tabular}{cc}
\includegraphics[width=8cm]{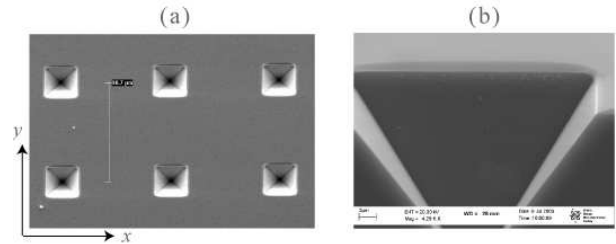}
\end{tabular}
\caption{SEM micrographs of the etched and gold-coated pyramids.
(a) top view showing pyramids in a rectangular array with a pitch
of $100\,\mu$m. (b) Cross-sectional view of a single pyramid. This
was obtained by cleaving the pyramid parallel to one of its edges.
The base of the pyramid has a side of length $30\,\mu$m,
corresponding to a perpendicular depth of $21.3\,\mu$m
}\label{SEM}
\end{center}
\end{figure}

We have fabricated 2-dimensional arrays of micro-mirrors in
silicon using a method that is simple, economical, and compatible
with MOEMS. We start with a $(100)$-oriented silicon wafer, coated
with a thin layer of oxide. Optical lithography is then used to
make square openings in the oxide, through which the silicon can
be etched. We use the anisotropic etchant potassium hydroxide
(KOH) at a concentration of $25\%$ by volume and a temperature of
$80\,{^\circ}$C. This attacks the $Si(100)$ plane more rapidly
than the $Si(111)$ plane, resulting in a pyramidal pit~\cite{Brendel}
bounded by
the four surfaces $(1,1,1)$, $(\bar{1},1,1)$, $(1,\bar{1},1)$ and
$(\bar{1},\bar{1},1)$. Typical resulting pyramids are shown in
Fig.\,\ref{SEM}. The $Si(111)$- faces of the pyramids are expected
to be extremely smooth because of the
 layer-by-layer etching mechanism involved\cite{{Moktadir},{Sato2}}. Atomic force
microscope measurements confirm this, giving an rms surface
roughness value of less than $0.5\,$nm
 for the uncoated pyramid faces. This makes them ideal as substrates for high-quality optical mirrors.
 After stripping the oxide mask away, a layer of gold of 100
nanometers thickness is applied to the silicon. Gold was chosen as it is a good reflector for
infrared light, but mirrors for other wavelengths can be made by using other metals or
multi-layer dielectric coatings. After sputtering, the surface roughness increases to $3\,$nm (rms).
 With this amount of roughness one can calculate that the scattering loss
of the specularly reflected intensity should be less than  $0.5\%$
in the near-infrared range~\cite{Beckmann}. By perfecting the coating
procedure it should be possible to decrease these losses to less
than $0.01\%$ in this region of the spectrum.

Fig.\,\ref{SEM}\,(a) shows a small section of the array viewed
under a scanning electron microscope after completion of the gold
coating. In this particular sample, the pyramids have a square
base whose sides are of $30\,\mu$m long. These are arranged in a
square lattice with a pitch of $100\,\mu$m. Both the etching and
the sputtering processes are standard techniques and can be
accurately controlled, guaranteeing that the fabrication results
are reproducible. Furthermore, large numbers of mirrors can be
packed at high density in a single batch. In the rest of this
paper we measure directly how the pyramids respond to light and we
show that they offer a promising new approach to fabricating
micro-mirrors for MOEMS.

\begin{figure}
\includegraphics[width=8cm]{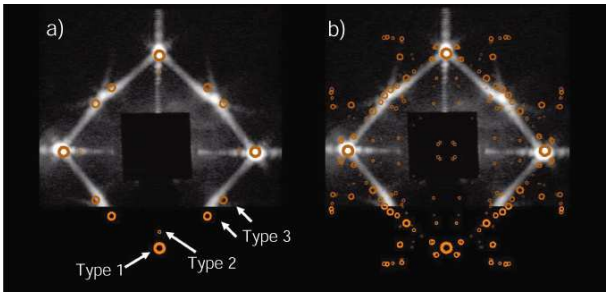}\\
\caption{Measured intensity distribution of reflected light, at a
distance of $7\,$cm from the array of pyramids, when it is
illuminated at normal incidence. A central bright spot, which is
caused by reflection from the region between pyramids, has been
covered to improve visibility of the light reflected from the
pyramids. The circles in (a) show the reflection pattern
expected for a perfect pyramid, while the circles in (b)
indicate the calculated reflection pattern for a pyramid with
rounded corners. Size indicates expected relative intensity.
}\label{reflectedIntensity}
\end{figure}

\begin{figure}
\includegraphics[width=8cm]{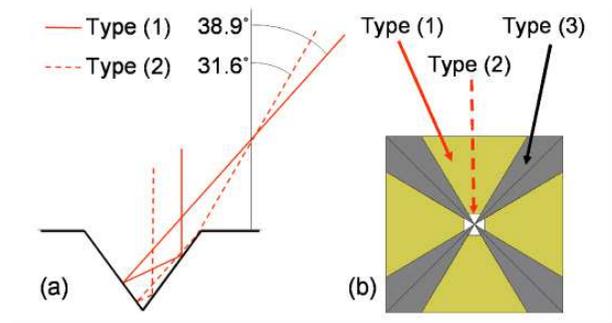}\\
\caption{(a) Cross section in the $x-z$ plane through a pyramid,
showing type (1) and type (2) trajectories.  These involve
reflections from mirrors on opposite sides of the pyramid. (b)
View of the entrance aperture of the pyramid, showing the regions
that produce type(1), type (2) and type (3) rays. }
\label{rays}
\end{figure}

 Our first test of the mirrors is to illuminate them with a collimated
$1\,$mm-diameter laser beam (wavelength $633\,$nm) propagating
along the $z$ axis, i.e. normal to the silicon surface and along
the symmetry axis of the pyramids. The sides of the pyramids
define $x$ and $y$ axes, as shown in Fig.\,\ref{SEM}.
Fig.\,\ref{reflectedIntensity}\,(a) shows the reflected pattern of
light observed on a screen $7\,$cm away from the mirrors. On this
we have drawn circles indicating the position of spots as expected
from a perfect pyramid. The three prominent spots at the corners
of the square are due to doubly reflected rays, which we classify
as type (1). These reflect from opposite faces of the pyramid, as
illustrated by the solid line in Fig.\,\ref{rays}\,(a). There
should be a fourth spot at the bottom of the photographs, but this
is blocked by a mount holding the beamsplitter through which the
array is illuminated.

 If the angle between opposite mirrors is
$\alpha$, the type (1) beams make an angle of $(\pi - 2 \alpha)$
with the $z$ axis. From the angles measured, we find that
$\alpha=(70.6\pm 0.7)^\circ$, in agreement with the expected angle
between opposing faces of $\arccos(1/3)=70.5^\circ$.
When the incident ray is close to the
apex of the pyramid (within $1.6\,\mu$m for a pyramid of
$30\,\mu$m base length), it is reflected twice by the first
mirror, as illustrated by the dashed line in Fig.\,\ref{rays}\,(a).
These rays, which we call type (2), should produce secondary spots
just inside the type (1) spots. However, the power in the type (2)
reflected beams is expected to be 100 times smaller because of the
small area from which they originate, as shown in
Fig.\,\ref{rays}\,(b).
Consequently it is not possible to identify the type (2)
beams clearly against the diffracted wings of the type
(1) beams.  Furthermore, there is a background of light
along the $x$- and $y$-axes caused by reflection from
rounded edges on the entrance aperture of the pyramid,
which can be seen in Fig.\,\ref{SEM}\,(b).

If a ray is incident near one of the corners of the pyramid, the
first reflection sends it off towards the opposite mirror, but is
is intercepted and deflected by the adjacent mirror before the
opposite mirror sends it out of the pyramid as a type (3) ray.
These rays make an angle of $31.5^\circ$ with the $z$ axis and
form double spots at azimuthal angles of $36.9^\circ, 53.1^\circ$,
etc. as shown in Fig.\,\ref{reflectedIntensity}\,(a). These spots
are less distinct than those of type (1) because the corners of
the pyramid are rounded, a feature that does not affect the type
(1) rays. Fig.\,\ref{reflectedIntensity}\,(b) shows the same
photographed reflection pattern, but here the superimposed circles
indicate the expected position and magnitude of spots reflected
from a pyramid with rounded corners. The roundness is included in
the ray-tracing model by four additional surfaces at each corner.
These are shaped to form approximate cone sections with radii of
$2.5\,\mu$m at the base and $0.825\,\mu$m at the apex of the
pyramid. The resulting reflection pattern closely matches the
photographed intensity distribution.

\begin{figure}[!htb]
\includegraphics[width=8cm]{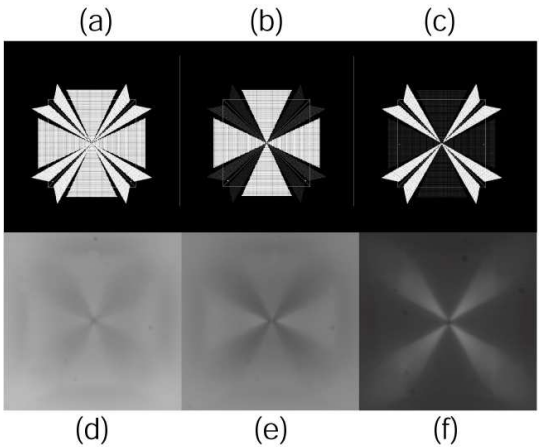}\\
\caption{Views of the vertex pyramidal mirror under an optical
microscope at $100\times$ magnification.  Top row: Raytracing
simulation. Bottom Row: Photographs. (a) and (d): without
polarisers; (b) and (e): parallel polariser and analyser; (c) and
(f): crossed polariser and analyser.} \label{microscopePics}
\end{figure}

The three types of rays described above also present different
characteristics when observed under illumination using polarised
light. Type (1) reflections leave the linear polarisation of the
light unchanged, whereas the type (3) reflections produce a
rotation of $\pm53^\circ$ or $\pm78^\circ$. This is investigated
in our second test of the mirrors, in which we examine them under
an optical microscope, illuminating them with white light once
again along the $z$ axis. Fig.\,\ref{microscopePics}\,(a) shows
the image calculated by raytracing for unpolarised light with the
microscope focussed in the plane of the apex of a perfect pyramid.
In this figure most of the area is bright. In
fig.\,\ref{microscopePics}\,(b) we show the expected image for
linearly polarised light, viewed through a parallel analyser,
which suppresses the type (3) contribution. This leads to a
reduction in the intensity of reflections from the corner region.
In Fig.\,\ref{microscopePics}\,(c), the analyser is crossed with
the polariser and only type (3) rays contribute, making the corner
region bright. The Intensity patterns observed in the laboratory
are shown in figures \,\ref{microscopePics}\,(d),
\,\ref{microscopePics}\,(e), and \,\ref{microscopePics}\,(f). They
correspond closely to the calculated distributions, indicating
that the pyramid reflects light as expected.

Knowing the types of reflected beams and their polarisations one
has the possibility of designing the optics to control the field
distribution within the pyramid. For example, appropriately angled
mirrors can be used to make closed resonant cavities containing
only type (1) rays of uniform polarisation. These can be either
standing-wave or ring cavities. Rays of type (3) give quite
different field distributions with a strong vertical standing wave
close to the axis of the pyramid, and horizontal standing waves
near the faces of the pyramid. Yet more elaborate field
distributions can be achieved by mixing the two. Potential
applications for the pyramids are in the areas of atom optics,
photonics and telecommunications. For example, by filling the pits
with ferroelectric material or liquid crystals and applying an
electric field, it should be possible to use the pyramids as fast
optical switches.

In summary, we have designed, fabricated and characterized a new
type of micro-mirror, produced by anisotropic etching through
square apertures on a silicon single crystal. Optical flatness is
achieved naturally through the layer-by-layer etching mechanism
involved. As an elementary component for optics, the micro-mirror
has a variety of possible applications in MOEMS devices, including
miniature optical switches and microcavities. Detailed experiments
and further theoretical analysis are currently under way to
develop these applications.

%\bibliography{pyramids}

\end{document}